\documentclass[conference]{IEEEtran}

\usepackage[T1]{fontenc}
\usepackage{graphicx}
\usepackage{amsmath,amssymb}
\usepackage{booktabs}
\usepackage{array}
\usepackage{tabularx}
\usepackage{algorithm}
\usepackage{algpseudocode}
\usepackage{amsthm}
\usepackage[hidelinks]{hyperref}

\hypersetup{hypertexnames=false}

\newtheorem{definition}{Definition}
\newtheorem{theorem}{Theorem}
\newtheorem{lemma}{Lemma}
\newtheorem{proposition}{Proposition}
\newtheorem{remark}{Remark}

\begin{document}

\title{Sound Constructive Refinement from Production Envelopes to Executable Manufacturing Schedules}

\author{
\IEEEauthorblockN{Runhao Liu\textsuperscript{1,*}, Gang Huang\textsuperscript{2}, Fei Ding\textsuperscript{3}, You Li\textsuperscript{4},\\
Guangzhen Yao\textsuperscript{5}, Yuxuan Wu\textsuperscript{6}, Jingcheng Shou\textsuperscript{7}, and Peng Zhang\textsuperscript{8}}
\IEEEauthorblockA{\textsuperscript{1}Polytechnic Institute, Zhejiang University, Hangzhou, 310015, China\\
\textsuperscript{2}School of Mechanical Engineering, Zhejiang University, Hangzhou, 310058, China\\
\textsuperscript{3}Alibaba Group, Beijing, 100102, China\\
\textsuperscript{4}School of Data Science and Engineering, Guangdong Polytechnic Normal University, Guangdong, 510665, China\\
\textsuperscript{5}College of Science, National University of Defense Technology, Changsha, 410073, China\\
\textsuperscript{6}School of Mathematical Sciences, Zhejiang University, Hangzhou, 310058, China\\
\textsuperscript{7}Department of Physics, Qiuzhen College, Zhejiang University, Hangzhou, 310058, China\\
\textsuperscript{8}School of Mathematical Sciences, Zhejiang University, Hangzhou, 310058, China\\
\textsuperscript{*}Corresponding author: runhaoliu@zju.edu.cn}
}

\maketitle

\begin{abstract}
Production planning and execution systems can interpret capacity,
compatibility, material, and timing commitments differently.  We treat this
semantic boundary as a constructive refinement problem in which a
rolling-horizon planner emits a machine-day \emph{production envelope}---an
explicit contract fixing production, order-fulfillment, mold-state, inventory,
outsourcing, and unmet-demand commitments---and a constructive procedure
refines it into order--machine--day allocations.  The procedure maintains
fulfillment, production, and capacity residuals.  A zero fulfillment residual
certifies that the allocation preserves the envelope's capacity,
compatibility, and delivery-window semantics; a nonzero residual records the
product, machine, mold, date, and quantity that blocked realization.  Under
divisible quantities, exact arithmetic, shared allocation semantics, no
additional admissibility filter, and a matched-progress condition, the
procedure discharges all fulfillment residuals; stricter execution policies
are diagnostic extensions rather than completeness tests.  We evaluate a
factory-provided anonymized instance with 37 product types, 150 orders, and
8.35 million requested units, a 5{,}400-run synthetic matrix, and dynamic
shocks.  Under explicitly declared accessory-supply, initial-inventory,
effective-capacity, due-window, divisible-allocation, and mold-state
assumptions, the industrial scenario reports \emph{100\% on-time delivery}, no
shell outsourcing, and 1.9--4.6\% changeover-driven capacity loss.  Under a
strict in-house accessory interpretation, the same data expose a \emph{load
ratio of 8.0763} and nonzero residuals.  The synthetic matrix locates the load
boundary of the guarantee.
\end{abstract}


\section{Introduction}
\label{sec:introduction}

High-mix manufacturing plans must stay executable under interacting discrete constraints: heterogeneous capacities, mold-dependent compatibility, setup losses, delivery windows, outsourcing, and synchronization between shell molding and accessory machining.  Enterprise Resource Planning
and Advanced Planning and Scheduling systems~\cite{rossit2019data,villalonga2021decision}
often rely on aggregate or rule-based logic that hides machine-level conflicts, so a plan can be acceptable at the planning level yet non-executable on the shop floor.

The difficulty is not only to optimize a production objective but to guarantee that a machine-day artifact refines into executable order-level allocations.  The planner reasons over daily quantities, inventory, and mold assignments; the scheduler must assign the committed quantities to compatible machine-day capacity exposed by the envelope.  This is a contract-preserving refinement problem~\cite{back1998refinement,alur1999reactive,benveniste2018contracts}: a valid envelope is a formal contract, and the scheduler synthesizes an implementation that preserves it.

\begin{figure*}[t]
\centering
\includegraphics[width=.80\textwidth]{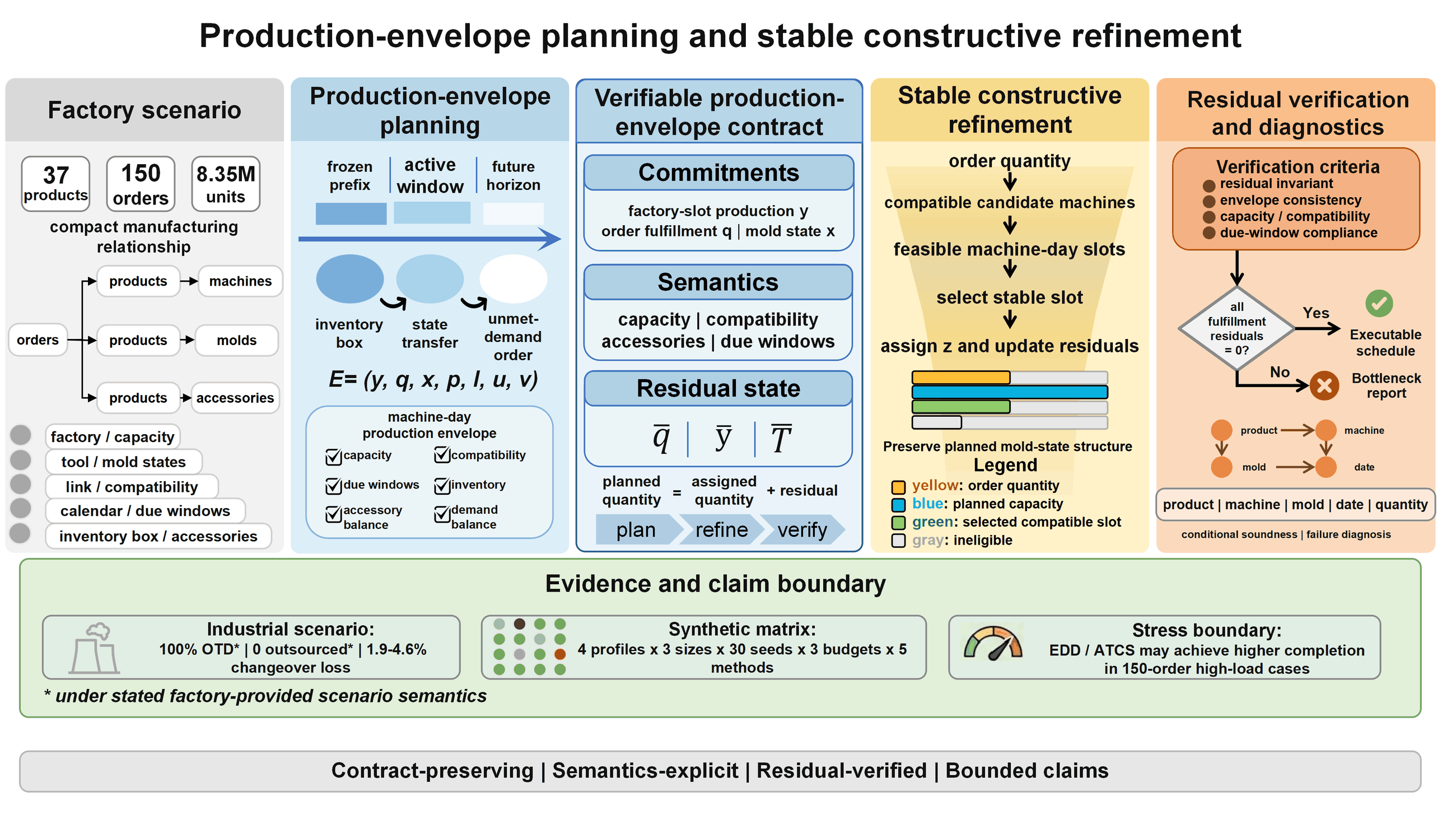}
\caption{Contract-preserving planning-to-execution interface.  The planner
emits a machine-day envelope; constructive refinement filters compatible
slots, updates residuals, and either releases an executable allocation or
returns a bottleneck record.  Industrial quantities shown here are conditional
on the accessory-supply semantics declared in Section~\ref{sec:evaluation}.}
\label{fig:overview_method}
\end{figure*}

We study high-mix scheduling as a \emph{production-envelope refinement problem}.  A rolling-horizon planner generates an envelope satisfying capacity, compatibility, inventory-balance, demand-accounting, and delivery-window constraints; a structure-aware scheduler (Figure~\ref{fig:overview_method}) then refines it under a one-mold-per-machine-per-day rule.  We establish residual invariants and prove that zero residual fulfillment implies the allocation is executable with respect to the envelope.

This differs from a conventional job-shop or dispatching study: existing models optimize a plan or a schedule, whereas the failure mode here occurs at their interface.  Each commitment is either synthesized into compatible capacity, covered by the envelope's outsourcing or unmet-demand variables, or returned as a traceable residual bottleneck.

The contributions are:
\begin{itemize}
\item We formalize high-mix scheduling as a \textbf{production-envelope refinement problem}, separating machine-day commitments from order-level allocation while retaining capacity, physical-mold exclusivity, product--mold--machine compatibility, delivery windows, outsourcing, setup-induced loss, and shared accessory capacity.
\item We give a \textbf{structure-aware constructive refinement} procedure that either synthesizes executable allocations or exposes the product, machine, mold, and date residuals that prevent refinement, and prove a \textbf{soundness theorem}: zero residual fulfillment implies the returned allocation preserves capacity feasibility, compatibility, and delivery-window compliance.  This is conditional soundness, not completeness for enriched execution models.
\item We \textbf{evaluate the interface} on an anonymized factory-provided scenario with 37 product types, 150 orders, and 8.35~million units, a 5{,}400-run synthetic matrix, a dynamic robustness suite, and a bounded solver reference, stating which assumptions the delivery result depends on.
\end{itemize}

\subsection{Related work}

Refinement calculus, abstract interpretation, interface theories, and model checking verify that lower-level implementations preserve higher-level specifications~\cite{back1998refinement,cousot1977abstract,henzinger2001interface,clarke1999model,baier2008principles,burch1990symbolic,necula1997proof}, and contract-based design treats components through assumptions and guarantees~\cite{benveniste2018contracts}; an envelope additionally binds continuous quantities, discrete mold states, and due-window commitments to a concrete allocation.
Manufacturing models and solver-backed synthesis address lot-sizing, tooling, and flexible job-shop problems with careful capacity and inventory treatment~\cite{pinedo2016scheduling,oujana2023mixed,sel2024energy,furlan2024matheuristic,rohaninejad2023integrated,solarlezama2008program,gulwani2017program,deMoura2008z3,biere2009handbook,baptiste2001constraint,laborie2018ibm,achterberg2009scip}, but they evaluate optimization quality or responsiveness; the residual interface instead makes a failed handoff observable as a structured mismatch~\cite{rossit2019data,villalonga2021decision,yuan2024machine,song2023flexible}.

Table~\ref{tab:related-positioning} summarizes the distinction.  Here,
$P$ denotes explicit production planning, $S$ detailed schedule synthesis,
$C$ contract or constraint verification, and $R$ refinement from an upstream
artifact; entries marked $P$ indicate partial coverage.
\begin{table}[t]
\centering
\caption{Positioning against related research streams.}
\label{tab:related-positioning}
\scriptsize
\begin{tabularx}{\linewidth}{@{}>{\raggedright\arraybackslash}p{.27\linewidth}*{4}{>{\centering\arraybackslash}p{.055\linewidth}}X@{}}
\toprule
Stream & P & S & C & R & Gap addressed here \\
\midrule
Refinement/model checking & N & N & Y & P & not specialized to quantity--machine envelopes \\
Contract-based design & N & N & Y & N & lacks production allocation semantics \\
CP/SMT scheduling & N & Y & P & N & solves schedules but rarely refines an upstream envelope \\
MILP production planning & Y & P & N & N & commits quantities without an execution certificate \\
Smart manufacturing dispatching & P & Y & N & N & improves responsiveness without contract-preservation proof \\
\textbf{This work} & \textbf{Y} & \textbf{Y} & \textbf{Y} & \textbf{Y} & \textbf{verifies envelope-to-allocation refinement} \\
\bottomrule
\end{tabularx}
\end{table}

\section{Production-Envelope Semantics}
\label{sec:semantics}

Let $M$ be machines, $F$ products, $K$ molds, $O$ orders, and $D=\{1,\ldots,H\}$ planning days, with $O_f=\{o\in O:f(o)=f\}$.  Order $o$ has product $f(o)$, demand $Q_o$, release day $r_o$, and due day $\ell_o$.  Mold $k$ can produce product $f$ only when $a_{f,k}=1$, and machine $m$ can host mold $k$ only when $b_{m,k}=1$.  Machine $m$ has a daily processing-time budget $T_{m,d}$, and product $f$ consumes $t_{m,f}$ time units per unit produced.  The static context is $\Pi=(M,F,K,O,D,f,Q,r,\ell,a,b,T,\tau)$, and all judgments below are relative to it.  Here $T_{m,d}=\widetilde T_{m,d}-S_{m,d}$ is the net budget after any setup loss already charged by the planning layer, which the refinement layer neither re-optimizes nor re-charges.

\begin{definition}[Production envelope]
\label{def:envelope}
A production envelope is a tuple
\[
E=(Y,q,x,p,I,u,v),
\]
where $Y_{m,f,d}\ge 0$ is the product allowance on machine $m$ and day $d$;
$q_{o,d}\ge0$ is the in-house fulfillment obligation for order $o$ on day $d$;
$x_{m,k,d}\in\{0,1\}$ is the installed mold state; $p_{f,d}$ and $I_{f,d}$ are
accessory production and inventory; and $u_o,v_o\ge0$ are outsourced and unmet
quantities.
\end{definition}

The allocation semantics needed by the refiner are mold exclusivity,
compatibility, capacity, order accounting, delivery windows, and
production-to-fulfillment consistency:
\begin{align}
&\sum_{k\in K} x_{m,k,d}\le 1, \label{eq:mold-exclusive}\\
&Y_{m,f,d}>0 \Rightarrow \exists k: x_{m,k,d}=1\wedge a_{f,k}=1\wedge b_{m,k}=1, \label{eq:compat}\\
&\sum_{f\in F}t_{m,f} Y_{m,f,d}\le T_{m,d}, \label{eq:capacity}\\
&\sum_{d=r_o}^{\ell_o} q_{o,d}+u_o+v_o=Q_o, \label{eq:demand}\\
&q_{o,d}=0\quad \text{for }d<r_o\text{ or }d>\ell_o, \label{eq:window}\\
&\sum_{m\in M} Y_{m,f,d}\ge \sum_{o\in O_f} q_{o,d}. \label{eq:prodfulfill}
\end{align}

\begin{definition}[Allocation and scenario validity]
\label{def:allocvalid}
An envelope satisfies \emph{allocation validity}, $E\models_{\Pi}\mathsf{AllocValid}$, when it satisfies Eqs.~\eqref{eq:mold-exclusive}--\eqref{eq:prodfulfill} and the declared domains.  The stronger predicate $E\models_{\Pi}\mathsf{ScenarioValid}_{\theta_A}$ additionally requires the planner-side accessory balance
\begin{align}
I_{f,d}&=I_{f,d-1}+p_{f,d}-\sum_{o\in O_f}q_{o,d}h_f, \qquad I_{f,d}\ge 0,
\label{eq:inv-balance}\\
\sum_{f\in F}\tau_f p_{f,d}&\le T^{\mathrm{CNC}}_d,
\label{eq:cnc-capacity}
\end{align}
with $h_f$ the equivalent accessory requirement per shell unit, $\tau_f$ the CNC time per equivalent accessory unit, and $T^{\mathrm{CNC}}_d$ the shared CNC budget.  Write $\mathcal A_{\theta_A}$ for their conjunction, with $\theta_A$ the accessory parameters $h$, $\tau$, $T^{\mathrm{CNC}}$, and $I_{\cdot,0}$.  Initial inventory and external accessory supply enter the right-hand side of Eq.~\eqref{eq:inv-balance} explicitly, and are not inferred from a successful shell schedule.  The planner evaluates $\mathcal A_{\theta_A}$ before refinement; it is \emph{not} certified by $z$.
\end{definition}

Eq.~\eqref{eq:inv-balance} is where the industrial result's governing accessory-supply assumption lives.

\begin{definition}[Executable allocation]
\label{def:exec}
An allocation $z$ assigns quantities $z_{o,m,d}\ge0$ to order--machine--day
triples.  It is executable with respect to $E$ when $z_{o,m,d}=0$ outside the
delivery window $[r_o,\ell_o]$; every positive entry admits a compatible
installed mold, $z_{o,m,d}>0\Rightarrow\exists k:x_{m,k,d}a_{f(o),k}b_{m,k}=1$;
aggregate processing time respects the budget,
$\sum_ot_{m,f(o)}z_{o,m,d}\le T_{m,d}$; allocations stay within the allowance,
$\sum_{o\in O_f}z_{o,m,d}\le Y_{m,f,d}$; and they never exceed the fulfillment
obligation, $\sum_m z_{o,m,d}\le q_{o,d}$.
\end{definition}

The base execution semantics are divisible at the machine-day level: several
orders may share one machine on a day provided their aggregate processing time
is within $T_{m,d}$.  A one-order-per-machine-day rule is an optional
restriction represented by $\Gamma$ in Section~\ref{sec:algorithm}.

\begin{definition}[Refinement]
\label{def:refinement}
Allocation $z$ refines envelope $E$, written $z\sqsubseteq_{\Pi} E$, when $z$
is executable and all in-house fulfillment commitments are discharged:
\[
\sum_{m\in M} z_{o,m,d}=q_{o,d}\quad \forall o,d.
\]
This excludes outsourced and unmet quantities from the executable in-house
allocation.
\end{definition}

For later use, define the allocation certificate
\[
\mathsf{Cert}_{\Pi}(E,z)\;\Longleftrightarrow\;
E\models_{\Pi}\mathsf{AllocValid}\ \wedge\ z\sqsubseteq_{\Pi}E.
\]
The optional planner-side predicate $\mathcal A_{\theta_A}$ is attached as a
separate witness when scenario validity is in scope.

\begin{definition}[Residual state]
\label{def:residual}
Given a partial allocation $z$, the residual state is
\begin{align*}
R(E,z)&=(\bar q,\bar Y,\bar T),\\
\bar q_{o,d}&=q_{o,d}-\sum_m z_{o,m,d},\\
\bar Y_{m,f,d}&=Y_{m,f,d}-\sum_{o\in O_f}z_{o,m,d},\\
\bar T_{m,d}&=T_{m,d}-\sum_ot_{m,f(o)}z_{o,m,d}.
\end{align*}
The predicate $R_q(E,z)=0$ denotes $\bar q_{o,d}=0$ for every order and day.
\end{definition}

Residuals serve two roles: they make the soundness argument local, and a
nonzero $R_q(E,z)$ names the commitment and semantic check that blocked
realization.

\paragraph{Refine-or-diagnose problem.}
Given $(\Pi,E,\Gamma)$, the interface must return either an allocation $z$
with $\mathsf{Cert}_{\Pi}(E,z)$, or a partial allocation together with a
nonempty tagged report $B$ identifying the precondition or residual blocking
the remaining fulfillment.

\section{Constructive Refinement Algorithm}
\label{sec:algorithm}

The refiner processes the envelope one day at a time.  For each active
order-day commitment it identifies compatible slots with positive planned
production, then removes slots without residual capacity.  To represent
execution rules absent from the envelope (batch, shift, or occupancy
restrictions), let $\Gamma(o,m,d;z,R)\in\{0,1\}$ be an optional
candidate-admissibility predicate.  The base refiner uses $\Gamma=\top$; an
enriched refiner only removes candidates and never enlarges the envelope.
The procedure selects a stable admissible slot, assigns the largest admissible
quantity, and updates the three residual families in their respective units.

More precisely, $\Gamma$ is \emph{deletion-only} when its candidate set at
every residual state is a subset of the base candidate set
$\mathcal S_{\top}=\{m\in\mathcal X:\bar T_{m,d}>0\}$.  The soundness result needs only this
monotonicity; it does not depend on what the extra rule represents.

Before Algorithm~\ref{alg:refinement}, the implementation evaluates
$\mathsf{AllocValid}$ and, when accessory semantics are in scope, consumes the
result of $\mathsf{ScenarioValid}_{\theta_A}$.  Violations are reported as
\textsc{InvalidWindow}, \textsc{InvalidCompatibility}, \textsc{InvalidCapacity},
\textsc{InvalidProductionBalance}, or an externally witnessed
\textsc{InvalidAccessoryScenario}.

\begin{algorithm}[t]
\caption{Contract-preserving constructive refinement}
\label{alg:refinement}
\begin{algorithmic}[1]
\Require $E\models_{\Pi}\mathsf{AllocValid}$, processing times $t_{m,f}>0$, admissibility filter $\Gamma$
\Ensure Allocation $z$ and residual report $B$
\State Initialize $z_{o,m,d}\gets 0$, $B\gets\emptyset$
\State $\bar q_{o,d}\gets q_{o,d}$, $\bar Y_{m,f,d}\gets Y_{m,f,d}$, $\bar T_{m,d}\gets T_{m,d}$
\For{$d\in D$}
  \For{orders $o$ with $\bar q_{o,d}>0$, sorted by due date}
    \While{$\bar q_{o,d}>0$}
      \State $\mathcal P\gets\{m:\bar Y_{m,f(o),d}>0\}$
      \If{$\mathcal P=\emptyset$}
        \State add \textsc{NoPlannedProduction} to $B$; \textbf{break}
      \EndIf
      \State $\mathcal X\gets\{m\in\mathcal P:m$ compatible with $x_{m,\cdot,d}\}$
      \If{$\mathcal X=\emptyset$}
        \State add \textsc{NoCompatibleProduction} to $B$; \textbf{break}
      \EndIf
      \State $\mathcal T\gets\{m\in\mathcal X:\bar T_{m,d}>0\}$
      \If{$\mathcal T=\emptyset$}
        \State add \textsc{NoResidualCapacity} to $B$; \textbf{break}
      \EndIf
      \State $\mathcal S_\Gamma\gets\{m\in\mathcal T:\Gamma(o,m,d;z,R)=1\}$
      \If{$\mathcal S_\Gamma=\emptyset$}
        \State add \textsc{NoAdmissibleSlot} to $B$; \textbf{break}
      \EndIf
      \State choose $m\in\mathcal S_\Gamma$ by the fixed lexicographic key (mold-state continuity, $\bar T_{m,d}$, $-\mathrm{id}(m)$)
      \State $\delta\gets\min\{\bar q_{o,d},\bar Y_{m,f(o),d},\bar T_{m,d}/t_{m,f(o)}\}$
      \If{$\delta\le 0$}
        \State add \textsc{NonProgress} to $B$; \textbf{break}
      \EndIf
      \State $z_{o,m,d}\gets z_{o,m,d}+\delta$
      \State $\bar q_{o,d}\gets\bar q_{o,d}-\delta$; $\bar Y_{m,f(o),d}\gets\bar Y_{m,f(o),d}-\delta$
      \State $\bar T_{m,d}\gets\bar T_{m,d}-t_{m,f(o)}\delta$
    \EndWhile
  \EndFor
\EndFor
\State \Return $(z,B)$
\end{algorithmic}
\end{algorithm}

The procedure is deliberately local: its purpose is not to solve a new global
scheduling problem after the plan has been fixed, but to implement the
envelope while preserving its mold-state structure.  At any step it subtracts
quantities only from residuals of a compatible planned machine-day slot that
passes $\Gamma$.  The lexicographic tie-break in line~22 makes the
construction deterministic for a fixed $(\Pi,E,\Gamma)$: the continuity score
comes from the declared mold-state sequence and ties are resolved by a fixed
machine identifier.  The planner may be a MILP model, a heuristic, or a
human-edited system; the theorem assumes only that the emitted envelope
satisfies the validity predicates of Section~\ref{sec:semantics}.  Each
successful assignment exhausts at least one positive residual entry and no
residual increases, so the procedure terminates in at most $N_q+N_Y+N_T$
assignments, and with precomputed order lists and constant-time queries the
running time is $O(|M|(N_q+N_Y+N_T)+|O||D|)$.

The returned bottleneck report is traceable to a product, date, installed mold
state, candidate machine set, and capacity conflict, which makes refinement
failure a diagnosable interface violation rather than an opaque scheduling
miss.  Table~\ref{tab:contract} summarizes the contract fields, the
corresponding refinement obligation, and the diagnostic tag emitted when the
obligation cannot be discharged.

\begin{table}[t]
\centering
\caption{Envelope fields as a contract, and the diagnostic tag emitted when
the obligation cannot be discharged.}
\label{tab:contract}
\scriptsize
\begin{tabularx}{\linewidth}{@{}lXX@{}}
\toprule
Field & Refinement obligation & Tag on failure \\
\midrule
$Y$ & Do not allocate more product than allowed on a machine-day & \textsc{NoPlannedProduction} \\
$q$ & Discharge in-house order quantities & residual after scan \\
$x$ & Use only compatible mold--machine--product triples & \textsc{NoCompatibleProduction} \\
$T$ & Keep aggregate processing time within capacity & \textsc{NoResidualCapacity} \\
$\Gamma$ & Respect optional execution restrictions & \textsc{NoAdmissibleSlot} \\
$I,p$ & Satisfy planner-side accessory balance $\mathcal A_{\theta_A}$ first & \textsc{InvalidAccessoryScenario} \\
$u,v$ & Keep outsourced and unmet demand outside the certificate & --- \\
\bottomrule
\end{tabularx}
\vspace{1pt}
\parbox{\linewidth}{\scriptsize A partial allocation with nonzero residuals is
never counted as a successful refinement, and an admissible slot that yields no
positive quantity adds \textsc{NonProgress}.  The certificate is the triple
$(\Pi,E,z)$, not $(E,z)$: the static compatibility, capacity, processing-time,
and delivery-window data in $\Pi$ are part of the checking context.}
\end{table}

\section{Soundness}
\label{sec:soundness}

The soundness theorem is conditional: zero checked fulfillment residuals imply
that the allocation satisfies the stated envelope semantics, and the theorem
alone asserts neither termination nor discharge.  The boundary matters because
practical execution models can add requirements beyond
Eqs.~\eqref{eq:mold-exclusive}--\eqref{eq:prodfulfill}---indivisible lots,
sequence-dependent setup times, shift calendars, or a
one-order-per-machine-day restriction can make an allocation-valid planning
artifact fail under a stricter execution model, and a disagreement about
capacity units can produce a residual without implying a physical shortage.
Accessory inconsistency is detected separately through
$\mathsf{ScenarioValid}_{\theta_A}$.  The proof therefore certifies positive
refinement results and structures negative outcomes; it neither proves global
infeasibility from a failed refinement nor replaces an exact feasibility solver
for enriched execution semantics.

\begin{lemma}[Residual preservation]
\label{lem:residual}
During Algorithm~\ref{alg:refinement}, the following identities are maintained
after every assignment:
\begin{align}
q_{o,d} &= \sum_{m\in M} z_{o,m,d}+\bar q_{o,d}, \label{eq:qres}\\
 Y_{m,f,d} &= \sum_{o\in O_f} z_{o,m,d}+\bar Y_{m,f,d}, \label{eq:Yres}\\
T_{m,d} &= \sum_{o\in O}t_{m,f(o)} z_{o,m,d}+\bar T_{m,d}. \label{eq:cres}
\end{align}
All residual terms remain nonnegative.
\end{lemma}

\begin{proof}
The algorithm initializes each residual to the corresponding envelope
commitment or capacity.  Every positive assignment $\delta$ is no larger than
the current fulfillment residual, production residual, and capacity residual
expressed in product units.  The update adds $\delta$ to $z_{o,m,d}$,
subtracts $\delta$ from $\bar q_{o,d}$ and $\bar Y_{m,f(o),d}$, and subtracts
$t_{m,f(o)}\delta$ from $\bar T_{m,d}$.  The identities are preserved and the
choice of $\delta$ maintains nonnegativity.
\end{proof}

\begin{theorem}[Soundness of constructive refinement]
\label{thm:soundness}
Let $E\models_{\Pi}\mathsf{AllocValid}$ and let $\Gamma$ be deletion-only.  If
Algorithm~\ref{alg:refinement} terminates with $\bar q_{o,d}=0$ for all $o,d$,
then the returned allocation $z$ satisfies $z\sqsubseteq_{\Pi}E$.
\end{theorem}

\begin{proof}
By Lemma~\ref{lem:residual}, zero fulfillment residuals imply
$\sum_m z_{o,m,d}=q_{o,d}$ for every order and day, so all in-house
fulfillment commitments are discharged.  The algorithm allocates only to slots
that pass the compatibility check against the installed mold state $x$;
$\Gamma$ can only remove such slots, so it cannot introduce an incompatible
assignment.  The residual capacity update ensures that cumulative processing
time never exceeds $T_{m,d}$.  Because $q_{o,d}$ is zero outside the delivery
window of order $o$, no positive allocation can occur outside that window.
Finally, Eq.~\eqref{eq:Yres} ensures that allocations do not exceed the
production envelope.  Hence $z$ is executable and discharges all in-house
fulfillment commitments.
\end{proof}

\begin{proposition}[Discharge under matched divisible semantics]
\label{prop:discharge}
Assume continuously divisible quantities, $t_{m,f}>0$, exact arithmetic, shared
$\mathsf{AllocValid}$ semantics, and $\Gamma=\top$.  In addition, assume the
\emph{matched-progress condition}: at every reachable residual state, each
active order--day pair has a compatible machine with
$\bar Y_{m,f,d}>0$ and $\bar T_{m,d}/t_{m,f}>0$.  Then
Algorithm~\ref{alg:refinement} discharges all fulfillment residuals,
$R_q(E,z)=0$.
\end{proposition}

\begin{proof}
Under the matched-progress condition, every nonzero fulfillment residual has a
nonempty candidate set and the minimum in line~20 is strictly positive.
Consequently, each successful iteration reduces at least one positive residual
(fulfillment, allowance, or capacity) to zero.  Residuals never increase and
there are finitely many entries, so the loop terminates after finitely many
positive assignments.  The only remaining exit is therefore
$\bar q_{o,d}=0$ for every active order--day pair, which is exactly
$R_q(E,z)=0$.
\end{proof}

\begin{remark}[Boundary of the guarantee]
Under continuous divisible quantities, shared $\mathsf{AllocValid}$ semantics,
exact arithmetic, and $\Gamma=\top$, the matched-progress condition gives the
discharge guarantee.  A nonzero residual in that base model indicates a
validity, implementation, or numerical inconsistency.  With an enriched
deletion-only $\Gamma$, indivisible lots, occupancy rules, or other shop-floor
semantics, a nonzero residual is instead a policy-dependent bottleneck.  In
neither case does failed refinement prove global infeasibility under an
alternative policy.
\end{remark}

\section{Implementation and Evaluation}
\label{sec:evaluation}

We implemented the planner and refiner as an Advanced Planning and Scheduling
prototype.  The evaluation asks which contract properties are checked when an
allocation is returned, what is exposed when a check fails, how the
industrial scenario behaves under its supplied semantics, how planned
mold-state structure is reflected in the allocation, and how delivery behaves
under perturbation.

\subsection{Instance and metrics}

The industrial instance comes from an anonymized smartphone protective-case
manufacturer.  It contains 37 product types, 150 orders, and 8.35~million
requested shell units over a 244-day horizon.  The equipment groups are eight
GT150 molding machines (4{,}800~units/day nominal), four GT130 molding machines
(3{,}000~units/day), and two SK750 CNC accessory machines
(2{,}000--4{,}000~units/day).  A molding-machine changeover incurs a five-hour
loss.  Accessory items are a heat ring (1.5~g aluminum) and a stand (12~g
aluminum); 4{,}255{,}000 rings and 5{,}755{,}000 stands are required, for
10{,}010{,}000 accessory pieces in total.

Three execution configurations are compared.  \emph{Scheme A}
(type-level abstraction) interprets the envelope only at the aggregate
machine-type level, exposing economic plausibility but not individual-machine
executability.  \emph{Scheme B} (unrestricted diagnostic execution) executes
the planned workload at machine level while allowing intra-day mold changes;
it is not the refinement relation certified by Theorem~\ref{thm:soundness}.
\emph{Scheme C} (stable constructive refinement) refines the envelope at
machine level under the one-mold-per-machine-per-day rule.  Earliest Due Date
(EDD) and Shortest Processing Time (SPT) dispatching are reported as
contract-free baselines.

Reporting uses on-time delivery
$\mathrm{OTD}=|O|^{-1}\sum_{o\in O}\mathbf 1[C_o\le\ell_o]$, with $C_o$ the
completion day and late orders counted by $C_o>\ell_o$; $\mathrm{LoadSigma}$, the
standard deviation of machine-level utilization
$\mathrm{Util}_m=\sum_d W_{m,d}/\sum_d T_{m,d}$, where $W_{m,d}$ is the
processing time the allocation assigns to machine $m$ on day $d$; the changeover ratio
$\mathrm{COR}=\sum_{m,d}S_{m,d}/\sum_{m,d}T_{m,d}$, the fraction of capacity
consumed by mold switching; and two accessory measures,
$\mathrm{SyncAcc}$ and the diagnostic gap $\mathrm{SyncGap}$, taken over
$\min(\mathrm{AccAvail}_d,\mathrm{AccNeed}_d)$ and
$|\mathrm{AccProd}_d-\mathrm{AccNeed}_d|$ against $\sum_d\mathrm{AccNeed}_d$,
where $\mathrm{SyncAcc}=1$ means availability fully covers demand while
$\mathrm{SyncGap}$ may still be nonzero.

\subsection{Results under the declared accessory semantics}

A result as strong as full delivery depends on the accessory balance of
Eq.~\eqref{eq:inv-balance}.  The source workbook records accessory geometry but no field for accessory inventory, external supply, or procurement cost, and accessory sourcing could not be confirmed with the partner plant.  We therefore \emph{declare} it as a labelled scenario assumption: accessory demand is met by external supply, closing the planner-side balance.  The
reported configuration further assumes the workbook's initial inventory and
due windows, the stated effective machine capacities, divisible order
quantities, and the planner's fixed mold-state sequence.  The base allocation
semantics permit multiple compatible orders to share a machine-day; the
one-mold-per-machine-per-day rule is the additional execution policy used by
Scheme~C.  These assumptions define the feasible region for the reported
industrial scenario.

Two consistency checks support the assumption: satisfying 10.01~million accessory pieces on the two SK750 units would require a cycle time of 4.2~s per piece, an order of magnitude below the 21.6--43.2~s implied by the workbook's own SK750 throughput; and the accessory material cost (1.51~million CNY at the workbook's aluminum price) is of the same order as the two SK750 assets (1.00~million CNY).

The third check is the interface itself.  Under the alternative reading, in which all accessory demand is produced in house, the instance requires 15{,}765{,}000 ring/stand time units against a documented horizon capacity of 1{,}952{,}000 --- an accessory load ratio of 8.0763.  The residual contract reports this as a \textbf{binding accessory residual}
while the GT150 and GT130 groups remain at 54\% and 75\% utilization, so the
mismatch is \textbf{localized to accessory synchronization} rather than presented as a
diffuse capacity shortage.  That is the diagnostic behavior the contract is
designed to provide, and it is why we report the sensitivity rather than
presenting the delivery result as unconditional.

Table~\ref{tab:industrial} reports the nominal configuration under these
declared factory-provided semantics.  Scheme~C \textbf{discharges the envelope}:
all 150 orders complete within their due windows with zero outsourced units
under the declared external-accessory-supply interpretation.  Scheme~B, which permits
intra-day mold changes, leaves 40 orders unserved and achieves 73.3\% OTD, so
additional shop-floor flexibility does not translate into additional
executable output.  EDD also reaches 100\% while SPT does not, but neither
contract-free rule exposes a relation to the upstream envelope.

\begin{table}[t]
\centering
\caption{Nominal-scenario validation of refinement strategies.  Industrial
values are conditional on the accessory-supply assumption declared above.}
\label{tab:industrial}
\scriptsize
\begin{tabular}{@{}lrrrrr@{}}
\toprule
Method & OTD & Late & SyncGap & LoadSigma & COR \\
\midrule
A (type-level) & 100.0 & 0 & 0.129 & 0.0775 & -- \\
B (machine, unrestricted) & 73.3 & 40 & 0.224 & 0.00135 & 0.0493 \\
C (one-mold-per-day) & 100.0 & 0 & 0.182 & 0.0161 & 0.0255 \\
EDD & 100.0 & 0 & 0.220 & 0.00212 & -- \\
SPT & 84.7 & 23 & 0.220 & 0.00474 & -- \\
\bottomrule
\end{tabular}
\vspace{1pt}
\parbox{\linewidth}{\scriptsize Reproduced from the shared workbook under the
conventions documented in the data note of Section~\ref{sec:boundaries}.  40
of 150 order records carry a client prefix that disagrees with the product
master, so the join convention is part of the reported configuration.}
\end{table}

\begin{figure}[t]
\centering
\includegraphics[width=\linewidth]{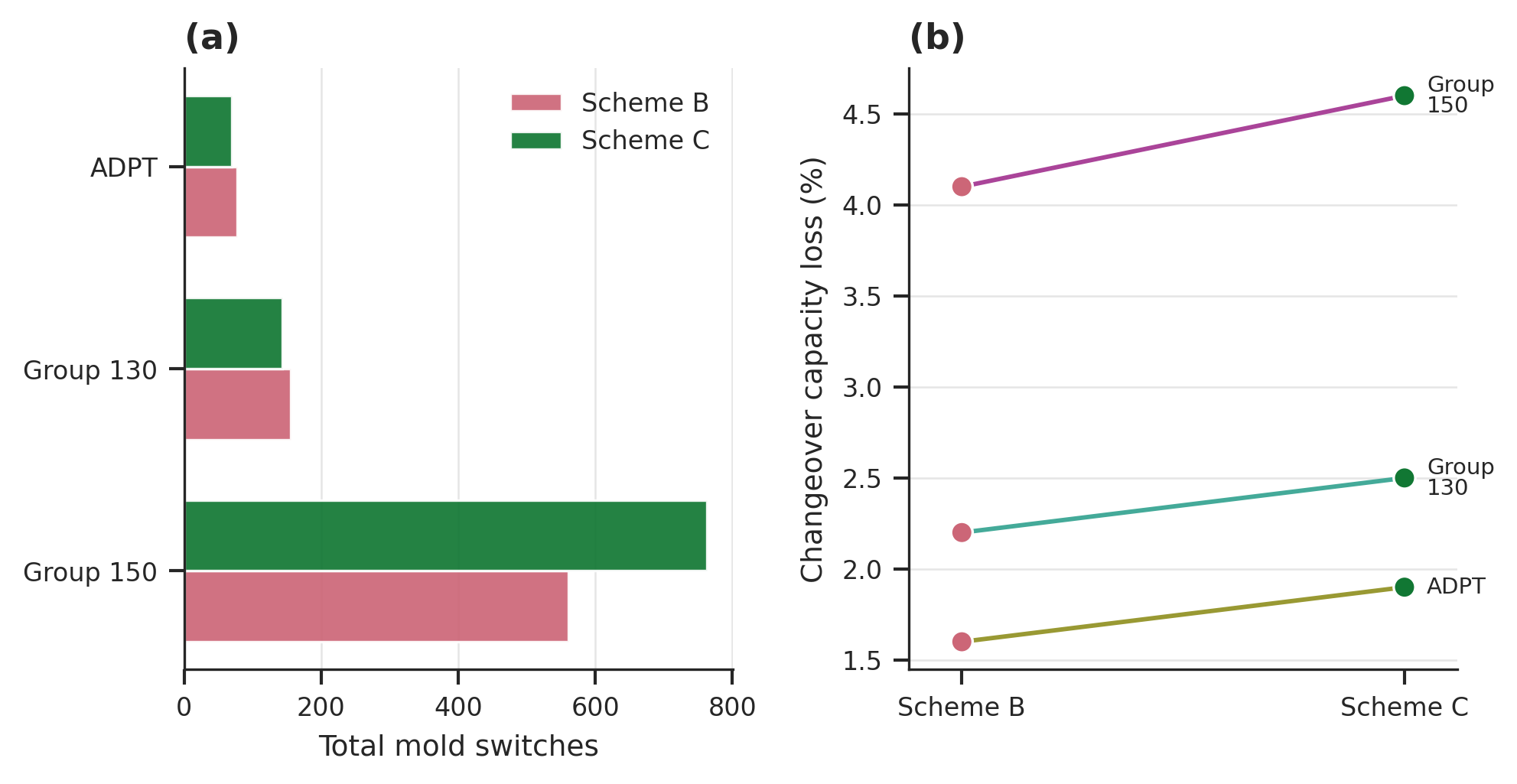}
\caption{Mold-switching and changeover-loss comparison by equipment group under
the declared industrial scenario.  Scheme~C trades a higher switch count for a
small, predictable loss because switching is concentrated at day boundaries.}
\label{fig:changeover}
\end{figure}

\begin{figure}[t]
\centering
\includegraphics[width=\linewidth]{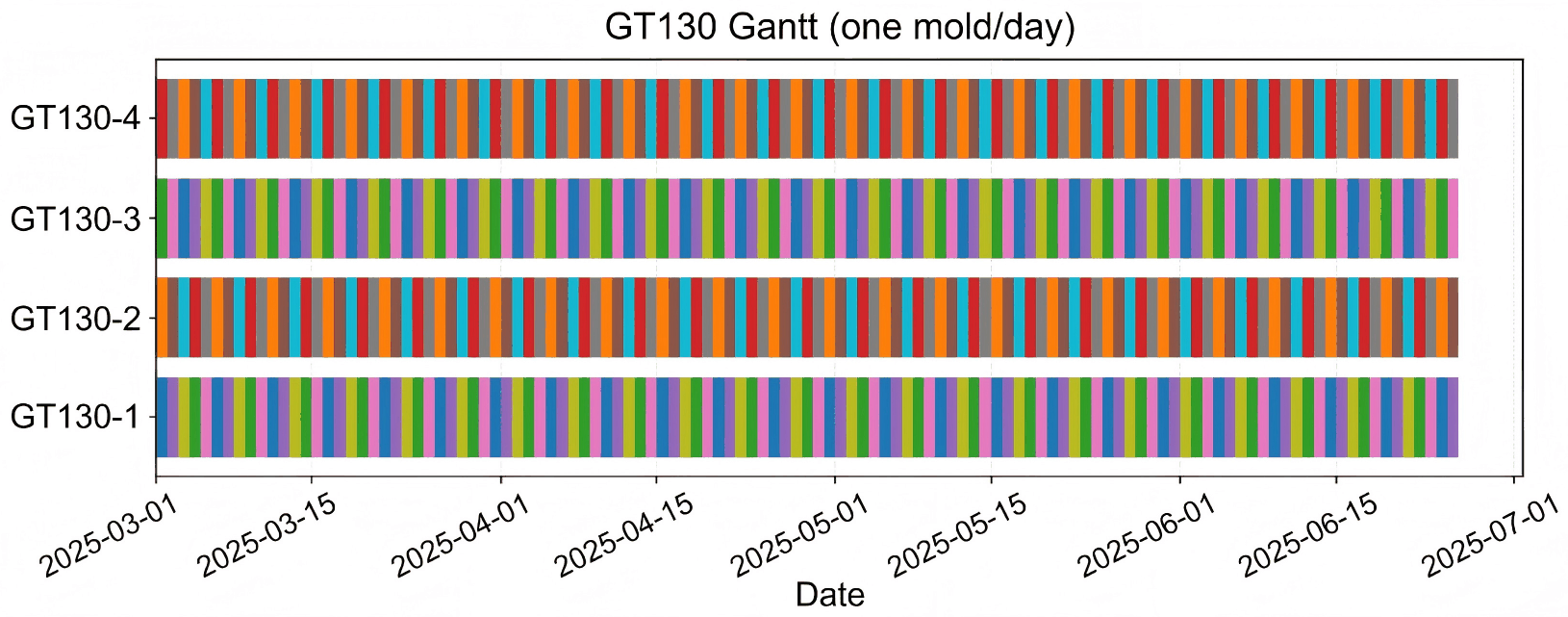}
\caption{Machine-level realization of the stable refinement on the GT130 group.
Each row is one machine; each colored block is one day under a single installed
mold, so a color change marks a mold change.}
\label{fig:gantt}
\end{figure}

The stability cost separates switch count from the capacity those switches
consume.  On the GT150 group Scheme~C performs more mold switches than
Scheme~B (763 against 561) but the induced capacity loss rises only from 4.1\%
to 4.6\%; on GT130 and CNC/ADPT the losses are 2.5\% and 1.9\% against 2.2\%
and 1.6\% for Scheme~B (Figure~\ref{fig:changeover}).  Scheme~C concentrates
mold changes at day boundaries, so the loss appears as a predictable reduction
of daily effective capacity rather than as fragmented productive intervals.
Figure~\ref{fig:gantt} shows the resulting machine-level structure.

\begin{figure}[t]
\centering
\includegraphics[width=\linewidth]{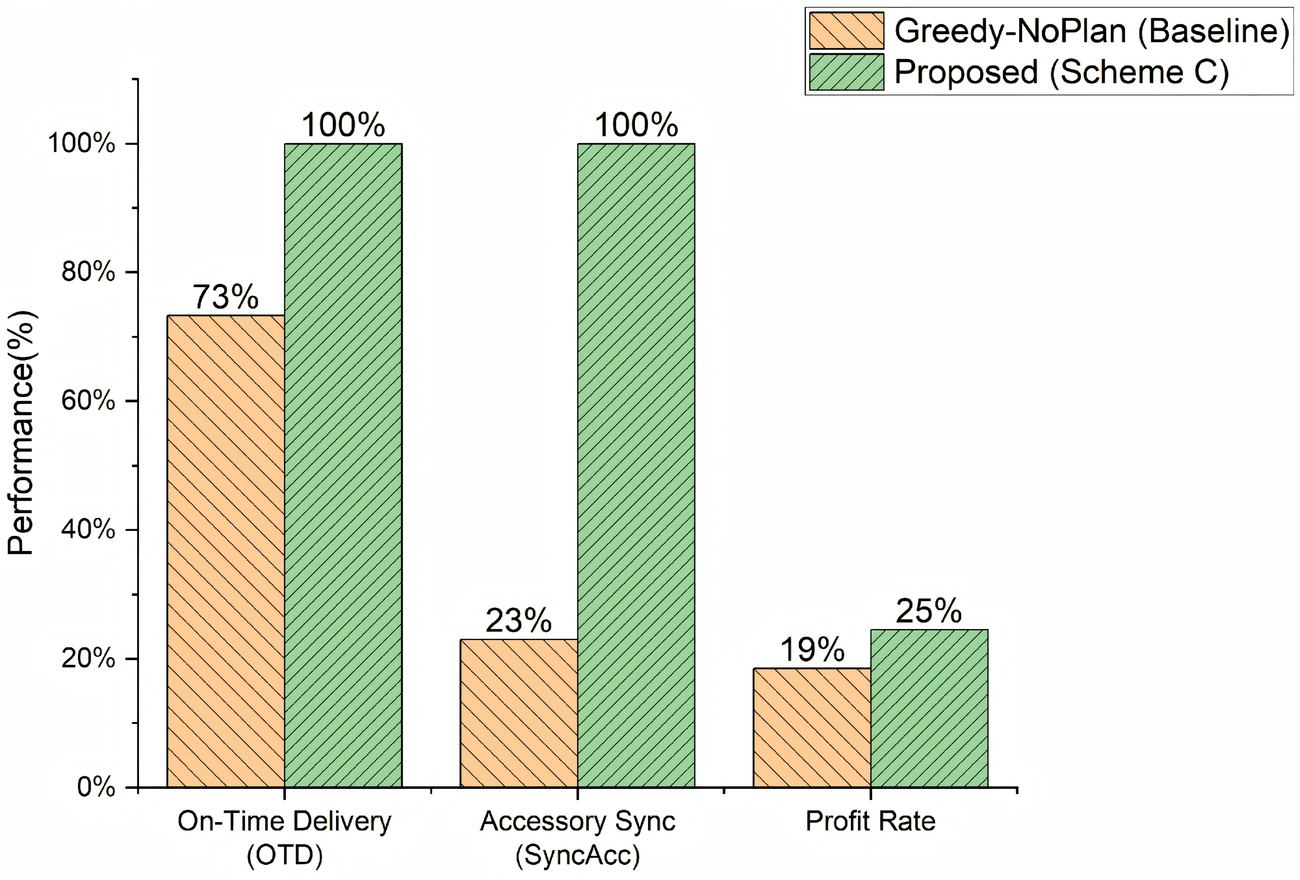}
\caption{Planning-contract ablation.  Removing the mid-term layer and executing
greedily at machine level collapses accessory synchronization accuracy from 1.00
to 0.23 and on-time delivery from 100\% to 73.3\%, so multi-stage
synchronization is not recovered without the envelope.}
\label{fig:ablation}
\end{figure}

Two further checks bound this result.  Scaling demand volume and
changeover-time coefficients to $\{0.8\times,1.0\times,1.2\times\}$ gives nine
perturbation scenarios; under the same declared accessory-supply semantics,
Scheme~C maintains 100\% OTD in all of them, whereas
EDD degrades to approximately 97\% at demand scale $1.2\times$ and SPT declines
to 82\%.  Removing the mid-term layer and executing greedily at machine level
instead collapses accessory synchronization accuracy from 1.00 to 0.23 and
on-time delivery to 73.3\% (Figure~\ref{fig:ablation}), the same shortfall
Table~\ref{tab:industrial} reports for the unrestricted machine-level scheme.
Both checks inherit the accessory-supply assumption of
Table~\ref{tab:industrial}.

\subsection{Synthetic evaluation: load and compatibility regimes}
\label{sec:synthetic}

The synthetic protocol is independent of the factory data and locates the
boundary of the contract-preservation claim.  It comprises four stress
profiles (balanced demand, tight due dates, high demand, and a compatibility
bottleneck), three order counts, 30 deterministic seeds, three solver budgets
(5, 30, and 60~s), and five methods: EDD, ATCS, rolling-horizon EDD (RH-EDD,
21-day lookahead with a seven-day frozen prefix), a rolling-horizon CP-SAT
optimizer (RH-CP-SAT, same windows), and the envelope-flow refinement of
Algorithm~\ref{alg:refinement}.  Every method shares one instance per seed and one validator, giving 5{,}400 observations.

Table~\ref{tab:matrix} reports the 60-second slice pooled over the four
profiles and 30 seeds.  At 30 orders the refinement matches the best completion
rate (99.5\%) while using roughly a third of the changeovers of every baseline
(11.6 against 34--40).  At 75 orders it stays within one point of RH-CP-SAT
while again changing fewer molds.  At 150 orders \textbf{the ordering inverts}: EDD and ATCS reach 58--59\% completion against 24\% for both envelope-respecting
methods, because a free dispatching rule may depart from the upstream envelope
while the refinement may not.

\begin{table*}[t]
\centering
\caption{Pooled synthetic results at the 60-second budget.  Each cell reports
completion mean $\pm$ SD (percent) / mean changeovers over 120 runs (four
profiles and 30 seeds).}
\label{tab:matrix}
\scriptsize
\begin{tabular}{@{}rccccc@{}}
\toprule
Orders & EDD & ATCS & RH-EDD & RH-CP-SAT & Envelope-flow \\
\midrule
30  & $98.3\pm5.97/39.9$  & $98.1\pm6.03/38.7$  & $98.3\pm5.97/39.8$  & $99.5\pm1.58/34.3$  & $99.5\pm1.58/11.6$ \\
75  & $90.7\pm12.46/127.9$ & $90.6\pm12.20/98.7$ & $90.7\pm12.46/127.3$ & $85.1\pm12.02/114.7$ & $86.3\pm12.19/100.2$ \\
150 & $59.3\pm21.57/310.7$ & $58.6\pm21.53/151.2$ & $59.2\pm21.49/311.0$ & $24.6\pm13.61/155.4$ & $24.4\pm12.24/152.4$ \\
\bottomrule
\end{tabular}
\end{table*}

\begin{figure}[t]
\centering
\includegraphics[width=\linewidth]{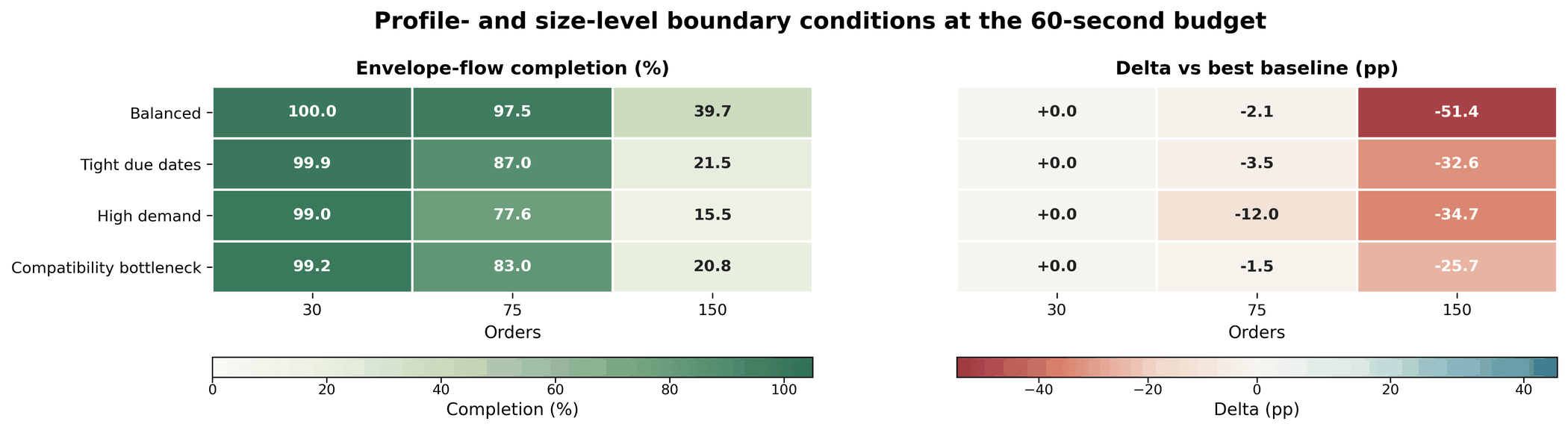}
\caption{Profile- and size-level completion boundary at the 60-second budget.
Contract validity and refinement soundness are established by the predicates
and the theorem; this heatmap is evidence about load regimes.}
\label{fig:matrix}
\end{figure}

This inversion is \textbf{a load boundary of the contribution} rather than a counterexample: under heavy load the question becomes whether the envelope was the right commitment, which re-planning must answer.  Figure~\ref{fig:matrix} gives the profile-level view, where the near-complete 30-order columns and the sharp 150-order degradation appear together.

\subsection{Dynamic robustness}

\begin{figure}[t]
\centering
\includegraphics[width=\linewidth]{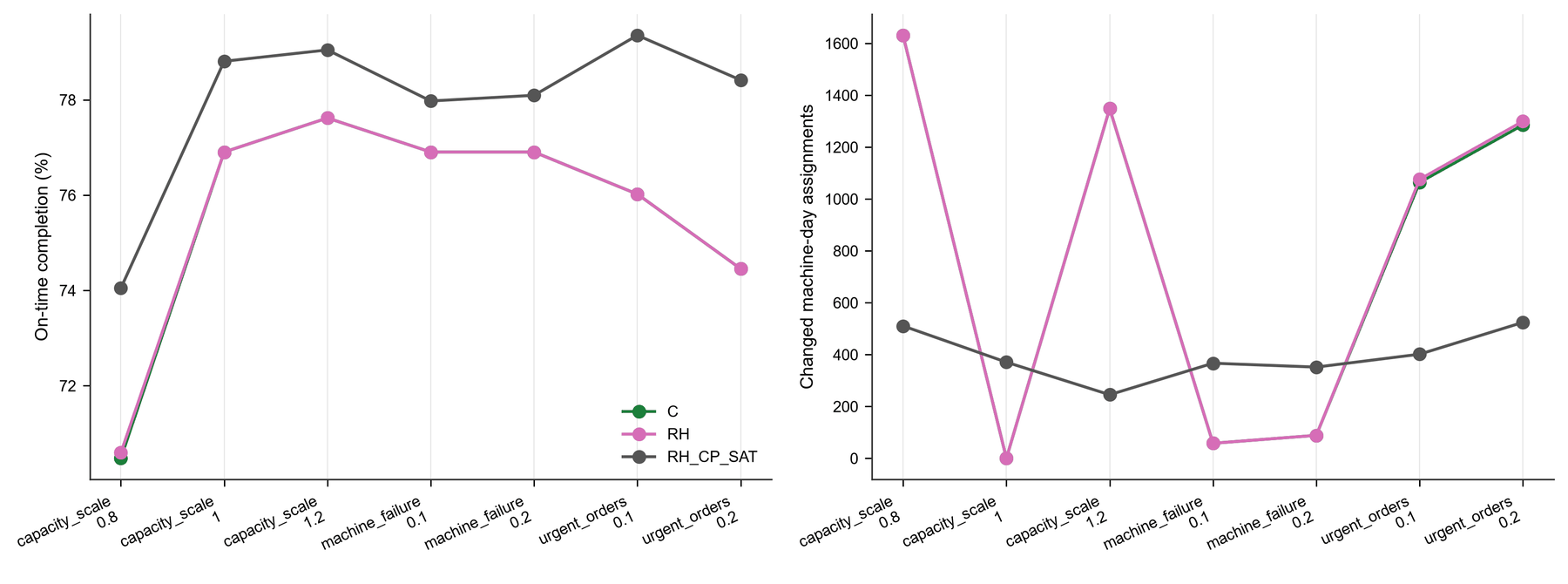}
\caption{Dynamic robustness on the high-demand profile across seven shock
levels.  Left: on-time completion.  Right: revised machine-day assignments.
C is the stable envelope refinement of Algorithm~\ref{alg:refinement}, RH a
rolling-horizon dispatching baseline, and RH-CP-SAT a rolling-horizon optimizer
that may re-plan inside its window.}
\label{fig:dynamic}
\end{figure}

A fixed envelope is vulnerable to shocks that invalidate its capacity or release assumptions.  We perturb the high-demand profile with a capacity scale of $0.8\times$, a machine failure removing 20\% of capacity, and an urgent-order arrival adding 20\% of demand.  Figure~\ref{fig:dynamic} reports completion and schedule nervousness --- the number of revised machine-day assignments --- for the stable refinement, a rolling-horizon dispatching baseline, and RH-CP-SAT, which may re-optimize inside its window.

Under machine failure the refinement is \textbf{markedly less nervous} than RH-CP-SAT (88.7 against 351.8 revised assignments), because it reallocates against frozen residuals instead of re-solving the window.  Under capacity and urgent-order shocks the ordering reverses: RH-CP-SAT completes one to four points more orders with fewer revisions, discarding the committed envelope.  \textbf{Severe shocks therefore call for re-planning rather than repair} at the refinement
layer, which is the response the bottleneck report is designed to trigger.

A small-instance comparison against an OR-Tools CP-SAT reference that additionally restricts each machine-day to at most one order returns 42 optimal, two feasible, and one unknown solution over 45 instances, with a mean completion gap of \textbf{0.0 percentage points} for both the stable refinement and EDD.  Its feasible set is \textbf{\emph{not} equivalent} to the refinement semantics, so it is completion context only.

\subsection{Failure modes and claim boundaries}
\label{sec:boundaries}

The main evaluation risk is to confuse optimization quality with refinement
soundness.  A dispatching rule may complete more units under heavy load because
it is free to reinterpret the envelope, which matters for operations management
but does not invalidate a theorem whose obligation is to preserve a fixed
contract; conversely, a zero-residual refinement proves only that the planner's
in-house commitments were implemented under the stated semantics.  Failed refinement is therefore \textbf{a first-class outcome}, and its record names the remaining quantity, product, date, and residual capacities, telling a planner whether the next action is re-planning,
capacity adjustment, mold-state revision, outsourcing, or data review.
Table~\ref{tab:records} shows the records the interface produced on the
industrial instance; each one is resolved to a \textbf{single binding state and a single planning action}, which is the property a completion rate does not provide.

\begin{table}[t]
\centering
\caption{Residual bottleneck records produced by the interface on the
industrial instance.  Each row is one emitted record.}
\label{tab:records}
\scriptsize
\setlength{\tabcolsep}{3pt}
\begin{tabularx}{\linewidth}{@{}lrlX@{}}
\toprule
Scope & Units & Binding state & Planning action \\
\midrule
GT130 case, day 42 & 18{,}500 & no stable compatible machine & reassign mold state \\
GT150 group & 11{,}200 & residual capacity exhausted & add capacity or outsource \\
Ring/stand CNC & 26{,}000 & accessory balance negative & verify inventory or supply \\
Order cluster, day 57 & 9{,}400 & residual near due date & re-plan earlier window \\
\bottomrule
\end{tabularx}
\end{table}

\emph{Data note.}  The shared workbook does not contain product identities
resolvable by exact string match for all orders: 40 of 150 order records carry
a client prefix that disagrees with the product master.  The product suffix is
unique in the master, and the loader maps each such record to its unique suffix
match and fails on ambiguity.  A literal string
join instead drops these records, which correspond to 26.7\% of orders and
2.22~million units, so the mapping convention changes the reported completion
rate materially.  We state the convention here because the comparison in
Table~\ref{tab:industrial} depends on it.

Several threats remain.  The scenario is factory-provided and anonymized, not
an independently audited plant model, and the accessory-supply assumption
behind Table~\ref{tab:industrial} is declared rather than verified; because the
policy preserves an upstream mold-state structure, unconstrained dispatching
may obtain better raw completion under high load.  The evidence consequently
supports sound, checkable refinement of a supplied envelope rather than general
scheduling superiority.

\section{Conclusion}
\label{sec:conclusion}

This paper formulates high-mix manufacturing execution as a sound constructive refinement problem: a production envelope is a contract between rolling planning and machine-day execution, and the refiner preserves it by assigning quantities only to compatible planned slots while maintaining residual invariants over fulfillment, production, and capacity.  The guarantee is intentionally bounded: zero fulfillment residuals imply an executable allocation, while remaining residuals are reported as bottlenecks.  The industrial scenario discharges the envelope under its declared accessory-supply semantics, the synthetic matrix locates the load boundary of the guarantee, and future work will extend the framework toward stochastic demand and exact infeasibility certificates.

\bibliographystyle{IEEEtran}
\bibliography{references}

\end{document}